# Cloud Service-Aware Location Update in Mobile Cloud Computing

Qi Qi[1,2,*], Yufei Cao[2]

*1. State Key Laboratory of Networking and Switching Technology, Beijing University of Posts and Telecommunications, Beijing, P.R. China, 100876.*

*2. EBUPT Information Technology Co., Ltd., Beijing, P.R. China, 100191*

** Corresponding author, Tel.: +86 13466759972, E-mail addresses: qiqi8266@bupt.edu.cn*

***Abstract*—Mobile devices are becoming the primary platforms for many users who always roam around when accessing the cloud computing services. From this, the cloud computing is integrated into the mobile environment by introducing a new paradigm, mobile cloud computing. In the context of mobile computing, the battery life of mobile device is limited, and it is important to balance the mobility performance and energy consumption. Fortunately, cloud services provide both opportunities and challenges for mobility management. Taking the activities of cloud services accessing into consideration, we propose a service-aware location update mechanism, which can detect the presence and location of the mobile device without traditional periodic registration update. Analytic model and simulation are developed to investigate the new mechanism. The results demonstrate that the service-aware location update management can reduce the location update times and handoff signaling, which can efficiently save power consumption for mobile devices.**

## I. INTRODUCTION

As mobile devices are becoming the primary platforms for many users who always roam around and access the cloud computing applications, the mobile computing and cloud computing have emerged as a new computing paradigm. Fernando et al. [1] summarize three mobile cloud computing concepts: (i) mobile device acts like a thin client connecting to the remote server through wireless network; (ii) mobile devices act both the resource providers and consumers and make up a peer-to-peer network; (iii) mobile device offloads its workload to a local 'cloudlet' comprised of several multi-core computers with connectivity to the remote cloud servers. By referring the above concepts, we discuss the first concept and consider the operation of mobile cloud computing through the core network IP Multimedia Subsystem (IMS). IMS is an overlay-architecture proposed by a large group of standardization organization (3GPP, 3GPP2, ETSI TISPAN), which combines heterogeneous access networks with IP-based services and targets to offer real-time IP multimedia applications over mobile networks [2].

On one hand, mobile cloud computing benefits from the IMS in means of increasing the system efficiency through the desired management mechanisms. Mobile cloud computing is used in the highly heterogeneous networks. Different mobile devices access to the cloud service through different access technologies such as Wideband Code Division Multiple Access (WCDMA), General Packet Radio Service (GPRS), (Worldwide Interoperability for Microwave Access) WiMAX, and Wireless Local Area Networks (WLAN). As a result, an issue of how to handle the wireless connectivity while satisfying the requirements of always-on connectivity, on-demand scalability of wireless connectivity, and the energy efficiency of mobile devices rises [3]. Fortunately, IMS provides service invocation, mobility management, QoS guarantee and security mechanism for mobile cloud services. The real-time telecommunication services in the cloud computing environment, which require significantly higher QoS levels and mobility management, can be provided by the IMS even during the access network change. On the other hand, the traditional telecommunication operators and service providers use cloud computing to realize the elastic resource utilization, especially for services shared by a large amount of users such as telecommunication service Voice over IP (VoIP), location, conferencing and messaging services, and Internet service image processing, natural language processing, multimedia search and sensor data applications [4]. To cope with high peak situations, telecommunication service infrastructures are usually over-provisioned, sometimes by an order of magnitude, and that leads to low resource utilization rate during typical times. Hence, hosting IMS application servers in the cloud enables pay-per-use cost models for traditional telecommunication service providers and reduces the risk of bad investments. Fig.1 shows the services in mobile cloud computing environment with a Cloud Service Monitor which is proposed to aware the mobile user's service behaviors.

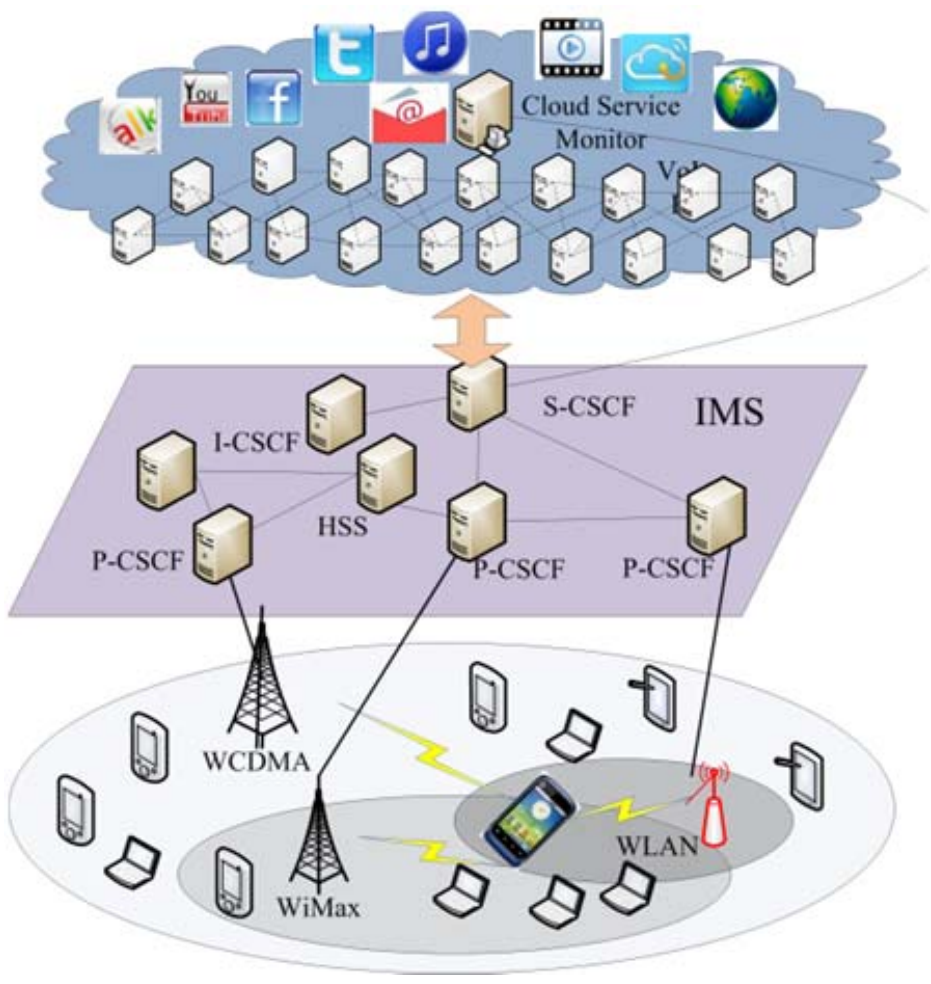

Fig.1. Mobile Cloud Computing

Mobile user may move from one access network to another or change its access technology. But the energy contained in its battery for a mobile device is finite, hence, one of the key issues encountered in a mobile cloud is the design of a power saved mobility management that support user mobility effectively. In this article, we focus on the new features that introduced by cloud services to the mobile computing environment and propose a complete solution to reduce energy consumption during their movement.

The remainder of this paper is structured as follows. Section II surveys the related work of mobile cloud computing and mobility management. Section III introduces the cloud service-aware location update solution to reduce energy consumption with original contribution presented in detail. Section IV models the proposed service-aware location update mechanisms. Section V analyzes the performance of the new mechanism. Section VI concludes this paper.

## II. RELATED WORK

The new concept of mobile cloud computing cannot be simply illustrated as merging mobile computing and cloud computing technologies which introduces many technology challenges.

As the battery lifetime of mobile devices is limited, and energy awareness for utilizing the cloud resource to reduce its energy consumption is a priority of designing applications on mobile phones. Rao et al. [7] propose a new approach for full-text keyword searches to retrieve content matching input keywords. For low energy, the keywords search is split into two subsets, such that one subset is processed by the mobile phone and another by the remote server. For the same purpose, Liu [8] et al. propose a mobile cloud system framework which can automatically perceive network conditions and smartly schedule data transmissions for different popular applications e.g., SNS and cloud storage for various multimedia content according to such conditions. This framework can lead to prolonged battery life and enhance user experience in mobile devices.

Ryu et al. [9] firstly work on the handoff mechanism for mobile cloud computing. The probability of predictive mode failure that distinguishes the predictive mode and the reactive mode of fast handovers for Mobile IPv6 (FMIPv6) are considered to optimize the handoff control in IP layer. The proposed scheme supports seamless handoff for various wireless technologies in cloud computing and reduces the overhead of traditional FMIPv6 protocol. Additionally, Choi et al. [10] propose a seamless connection handoff mechanism between different access networks in transport layer. By setting a checkpoint, this mechanism clear communication channel before switching the socket of connection from the old network to the new network. By this way, it guarantees that there are no transit data on the communication channel at the moment of connection handoff. Vrat et al. [11] propose a model to enhance the mobility services in cloud by utilizing the concept of Hierarchical Mobile IPv6 (HMIPv6) in coexistent network. It reduces the burden on existing IPv4 addresses and enhances mobility as a service in cloud computing. Tuncer et al. [16] presents a Virtual Mobility Domains on their Floating Cloud Tiered Internetworking model. The Virtual Mobility Domains supports both macro and micro mobility by leveraging a new tiered addressing, a network cloud concept, and a unique packet forwarding scheme. But non-IP addressing and classic routing protocols is not suitable for the mobile device in current mobile cloud computing environment.

For 2G and 3G nework, a normal location area update caused by a mobile device changing its location and a periodic location area update schemes are adopted in UMTS and GSM network, and dynamic periodic location area update is proposed in [12]. Based on this, the dynamic and optimal combination of periodic and normal location area update are provided in [13][14][15] with comparison to the cost of location update among these mechanisms.

As mentioned above, the existing research on mobile cloud computing for reducing the power consumption is offloading application in mobile device to remote server. The location update is indeed a forgotten procedure for mobile devices, which can be optimized for power saving. And the users' service behaviors in cloud computing environment such as service accessing have not been considered into location update.

## III. CLOUD SERVICE-AWARE LOCATION UPDATE

### *A. Motivation*

IMS obtains the location and state information of mobile devices through registration process. A binding is created by the S-CSCF based on initial registration between the public user identify and the IP address of mobile device. The S-CSCF and P-CSCF keep user's registration status including the timer that is indicated by the "expires" parameter in 'Contact' header of REGISTER message. The subsequent re-registrations are performed periodically to update the binding and refresh the timer [2].

In order to refresh the exiting registration or notify the network that the capability of mobile device has been changed, the mobile device should initiate re-registration process. The re-registration is triggered by two situations: Periodic Re-Registration (PRR) is used to periodically report the "active" state of mobile device to the network for avoiding expiry of the agreed registration timer between mobile device and network; 2) Re-Registration for Change Capabilities (CCRR) is initiated when the capabilities of the mobile device have changed. Typically it is the case that, as the mobile device moves and connects to another network entry point, registration procedure is performed once more to update registration timer in IMS. The goal of PRR is to detect if the mobile device is still registered to the network (the network will initiate de-registration procedure as the registration timer is timeout) while the goal of CCRR is to inform the network when and where the locations and capabilities of mobile device have been changed. The former is triggered by the timer periodically and the latter is triggered when capability parameters change. Both the PRR and CCRR process need to refresh the user's registration timer so that the mobile device can be located or it can initiate a new session.

For the IMS, if the network receives a session invited by the mobile device or the mobile device can response a session successfully, the IMS can determine the mobile device is still in the registration status. Different from traditional telecommunication network, the powerful mobile phones in the mobile cloud computing environment access kinds of Internet services, such as weather, instant messaging, social network, Internet TV, online gaming, document processing and other remote web applications. The behaviors of accessing the cloud services also indicate that the mobile device is in the registration status. Therefore, taking use of the session establishment and the cloud service accessing to refresh the registration timer can reduce the PRR times, which can save the power of the mobile device caused by the periodically re-registration.

### *B. Signaling procedure of cloud service-aware location update*

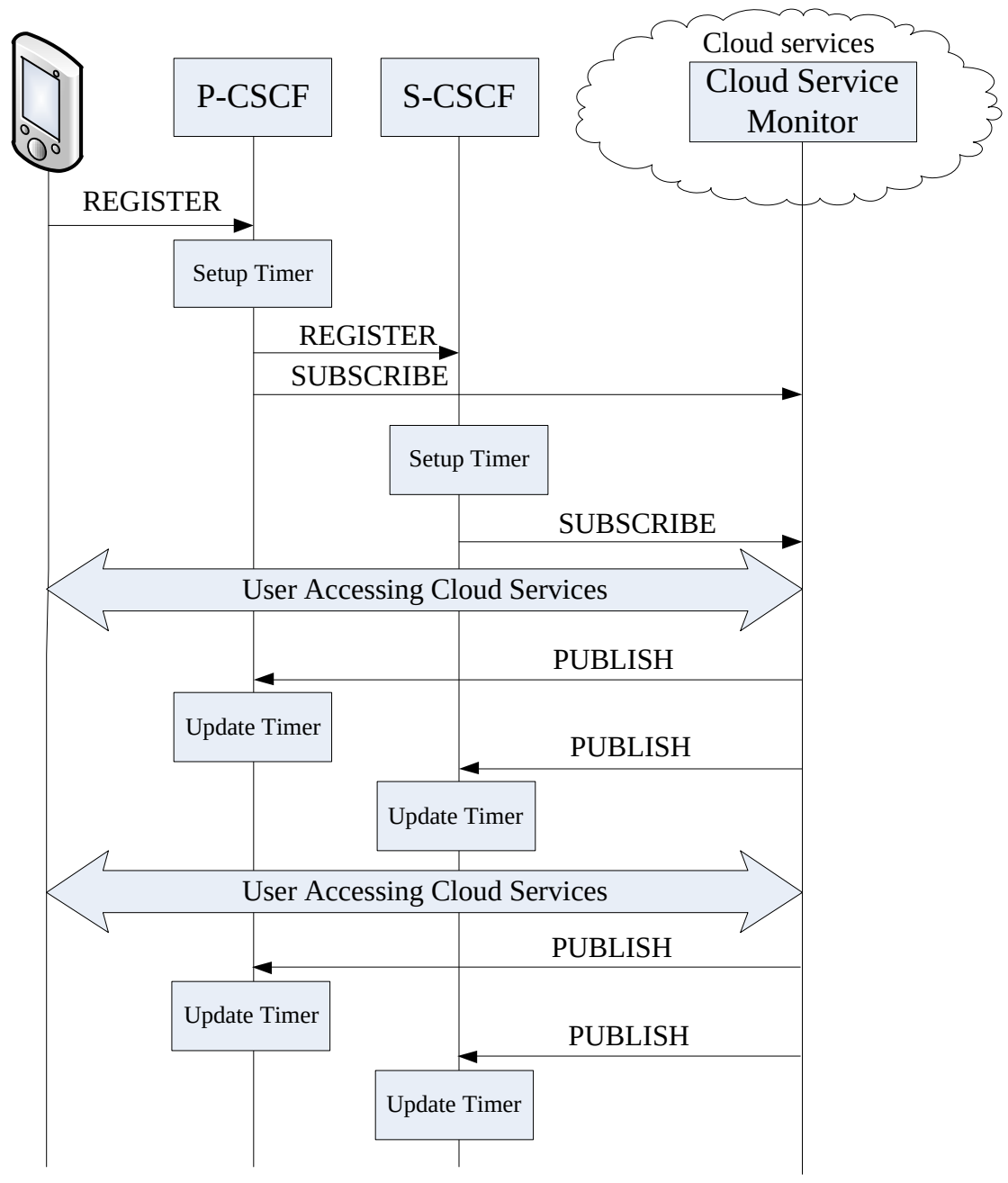

Fig.2. The Publish Signaling for location aware by cloud services

Besides SIP protocol, the cloud services may use other application layer protocol, such as HTTP. But the S-CSCF and P-CSCF can only process SIP message, so the IMS cannot obtain the cloud service accessing events derectly. By taking use of the event when the mobile device accesses the cloud service, the Cloud Service Monitor perceiving the user's behavior during its online time and obtain the information of mobile's devices status. The SUBSCRIBE/PUBLISH method is used to notify IMS the active state of the mobile device, depicted in Fig.2. The S-CSCF and P-CSCF subscribe the status of registered mobile devices from the Cloud Service Monitor.

For avoiding the large network signaling traffic between the Cloud Service Monitor and the S-CSCF and reducing the load of the cloud services, the Cloud Service Monitor capture the user’s behavior from the cloud services in an interval, and publishes its information to the S-CSCF and P-CSCF.

### *C. Algorithm of cloud service-aware location update*

We define user’s behavior checkpoint as an action to inform the IMS whether the mobile is attached or not. The checkpoint event includes a session establishment request, or CCRR event, or the publish message from the Cloud Service Monitor. Such an event results in checkpoint action.

1) Incoming or Outgoing Session (IOS) event: IMS receives the session initiated by the mobile device or the mobile device receives a session request from the IMS and responses successfully, such as a VoIP session setup request, or session request for a streaming service in the cloud;

2) Cloud Service Access (CSA) event: the Cloud Service Monitor publishes the access events to the S-CSCF and P-CSCF when the mobile device accesses the cloud service. For example, a user may logon the twitter and watch his friends’ content.

3) CCRR event: when the capabilities of the mobile device have changed, CCRR event occurs;

4) PRR event: the periodic re-registration to the IMS network.

The dynamic location update algorithm is as follows.

```
1.    If IOS event occurs
        If IMS receives session request from the mobile
      device
            IMS reset registration timer;
        End If
        If mobile device receives the response from IMS
            Mobile device reset registration timer;
        End If
      End If
2.    If CSA event occurs
        Cloud Service Monitor send PUBLISH message
        to S-CSCF and P-CSCF;
       Cloud Service Monitor embedded the timer in
     service response;
      End If
      If S-CSCF receives PUBLISH message from Cloud
     Service Monitor & mobile device's ID  in PUBLISH
     = mobile device ID in S-CSCF
          IMS reset registration timer;
      End If
      If Mobile device receives service response
containing a timer
            Mobile device reset registration timer;
       End If
      End If
3.    If CCRR event occurs
     Mobile device initiate re-registration;
     If IMS receives the re-registration request
       IMS reset registration timer;
     End If
     If mobile device receives response
       Mobile device reset registration timer;
     End If
       End If
4.    If registration timer =0 // PRR event occurs
          Mobile device initiate re-registration;
          If IMS receives the re-registration request
       IMS reset registration timer;
     End If
     If mobile device receives response
       Mobile device reset registration timer;
     End If
       End If
```

When the mobile device is disconnected from the network due to abnormal reasons, such as battery removal, subscriber moving out of the service area, and so on, the IMS detects abnormal detach of the mobile device in one of the following two cases, depicted in Fig. 3 (a). (1)The mobile device disconnected at t2 and the next re-registration for timer expires (t3) occurs before the arrival of the incoming session (t4). In this case, the network can detect expiration of the timer and considers the mobile device detached. The next incoming session will not be delivered. (2) The next incoming session occurs (t8) before the timer expires (t9), i.e. the timer update is not timely. In this case, the IMS attempts to deliver the next incoming session to the mobile but fails. After failure of the session setup, the IMS considers that the mobile device is detached and will disable future session terminations and registered timer.

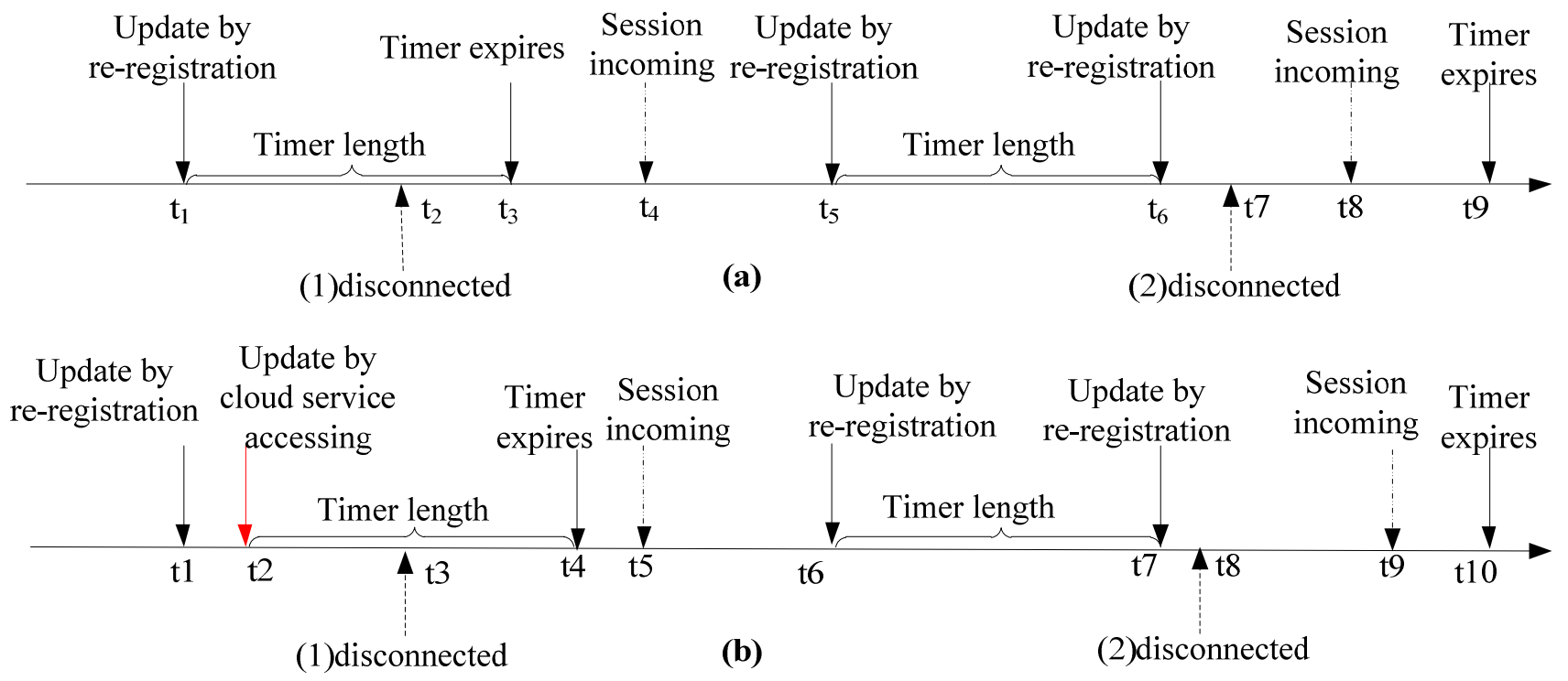

Fig. 3 The timer sequence of location update and mobile disconnected detecting

Then the session establishment, the cloud service accessing and the capability changing event are taking into updating the timer. When these events occur, the location is updated passively (t2) and the periodic re-registration event can be put off (t4). But when there are no above events, the periodic re-registration recovers (t6 and t7). From Fig. 3 (b), we can see the putting off the location update of re-registration has no effect of the IMS detecting disconnected mobile devices.

## IV. ANALYTICAL MODEL

In this section a formula will be derived for analyzing the power cost of location update. The PRR and CCRR event is sending REGISTER message to IMS, and IOS event is sending or receiving the INVITE message. The energy consumed in these processes is dependent on the used modulation and coding scheme and the exact number of sub-frames required for the process. However, for the purpose of analysis, we consider this power cost to be directly proportional to the size of the messages exchanged [17]. Thus, the power cost of the mobile device for processing update date point event $e$ is denoted by $c_e$ = L (message size). The CSA event is accomplished by Cloud Service Monitor notifying the S-CSCF when the mobile device accesses the cloud service. In this procedure, the mobile device only needs to update its timer by the application within its operation system. The energy consumption of this procedure is far less than sending or receiving a SIP message.

Assume the arrivals of IOS, CCRR and CSA events follow Poisson distribution and independent with each other. When IMS receives the update event, it refreshes the timer, which is called update check point. As the sum of several Poisson distribution is still a Poisson distribution, the arrival rate of update check point is

$$\lambda = \sum_{e=\{IOS,CCRR,CSA\}} \lambda_e \quad (1)$$

For each update check point, the number of a certain event is $\lambda_e / \lambda$. Then the power cost of the events for one update check point is

$$Power_e = c_e(\lambda_e / \lambda), \text{ where e=\{IOS, CCRR, CSA\}} \quad (2)$$

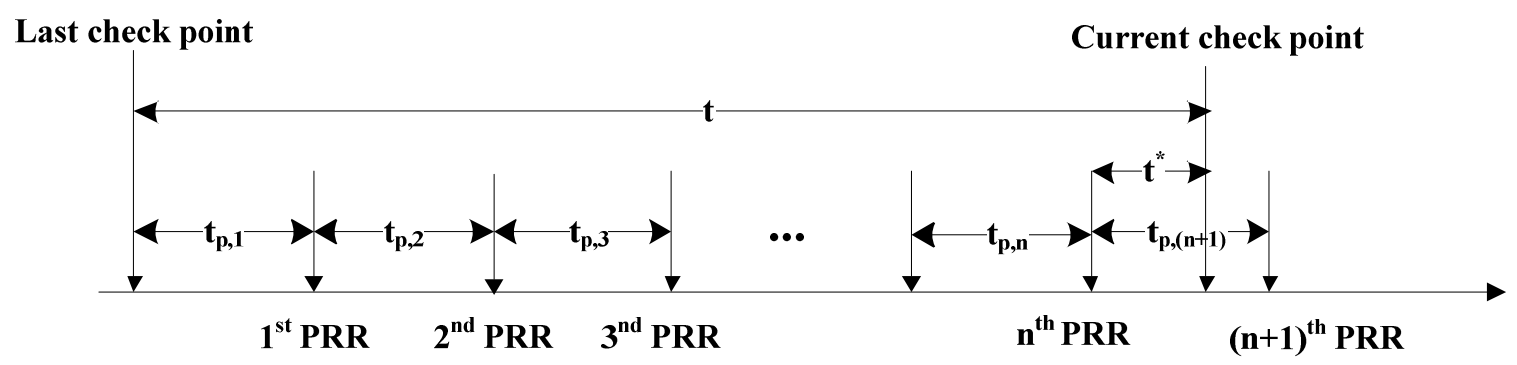

Fig 4. The number of PRR event during two update check points

IMS update event only contains CCRR, which means the update check point is receiving CCRR. In Fig.4, the time interval of update check point follows exponential distribution with a mean of $\lambda_{CCRR}$, and its probability density function is: $f(\lambda_{CCRR}) = \lambda_{CCRR} e^{-\lambda_{CCRR} t}$. Assume $N_{PRR}^{IMS}$ is the number of PRR event times during two update check points for IMS location update. As $t_{p,1} = t_{p,2} = ... = t_{p,n} = t_{p,(n+1)} = t_p$, $t = N_{PRR}^{IMS} t_p + t^*$. Then we can deduce the probability of $N_{PRR}^{IMS} = n$ during two update check points as:

$$P(N_{PRR}^{IMS} = n) = P(nt_p < t < (n+1)t_p) = \int_{nt_p}^{(n+1)t_p} \lambda_{CCRR} e^{-\lambda_{CCRR} t} dt = e^{-\lambda_{CCRR} n t_p} (1 - e^{-\lambda_{CCRR} n t_p}) \quad (3)$$

And the expectation of $N_{PRR}^{IMS}$ is:

$$E(N_{PRR}^{IMS}) = \sum_{n=1}^{\infty} nP(N_{PRR}^{IMS} = n) = \sum_{n=1}^{\infty} e^{-\lambda_{CCRR} n t_p}(1 - e^{\lambda_{CCRR} n t_p}) = \frac{1}{e^{\lambda_{CCRR} t_p} - 1} \tag{4}$$

Hence, the power cost of PRR event is

$$Power_{PRR} = c_{PRR} E(N_{PRR}^{IMS}) \tag{5}$$

The total power cost of the IMS location update and session establishment is

$$TPower_{IMS} = Power_{CCRR} + Power_{IOS} + Power_{PRR} = c_{CCRR}\frac{\lambda_{CCRR}}{\lambda} + c_{IOS}\frac{\lambda_{IOS}}{\lambda} + c_{PRR}\frac{1}{e^{\lambda_{CCRR} t_p} - 1} \tag{6}$$

In the service-aware location update algorithm, the location update events include CCRR events, IOS and CSA events. The update check point is the occurrence of CCRR, IOS or CSA event. The time interval of update check point *t* follows exponential distribution with a mean of $\lambda$, and its probability density function is: $f(\lambda) = \lambda e^{-\lambda t}$.

Taking use of the same method of modeling the standard IMS location update, we can also obtain the probability of $N_{PRR}^{Cloud} = n$ and the expectation of $N_{PRR}^{Cloud}$ during two update check points in cloud service-aware location update:

$$P(N_{PRR}^{Cloud} = n) = e^{-\lambda n t_p}(1 - e^{-\lambda n t_p}) \tag{7}$$

$$E(N_{PRR}^{Cloud}) = \frac{1}{e^{\lambda t_p} - 1} \tag{8}$$

Then power cost of PRR event for service-aware location update is

$$Power_{PRR} = c_{PRR} E(N_{PRR}^{Cloud}) \tag{9}$$

As the CSA event only need to update the timer in mobile device, the total location update power cost for service-aware location update includes the cost of CCRR, IOS and PRR events:

$$TPower_{Cloud} = Power_{CCRR} + Power_{IOS} + Power_{PRR} = c_{CCRR}\frac{\lambda_{CCRR}}{\lambda} + c_{IOS}\frac{\lambda_{IOS}}{\lambda} + c_{PRR}\frac{1}{e^{\lambda t_p} - 1} \tag{10}$$

## V. SIMULATION AND PERFORMANCE ANALYSIS

### A. Simulation

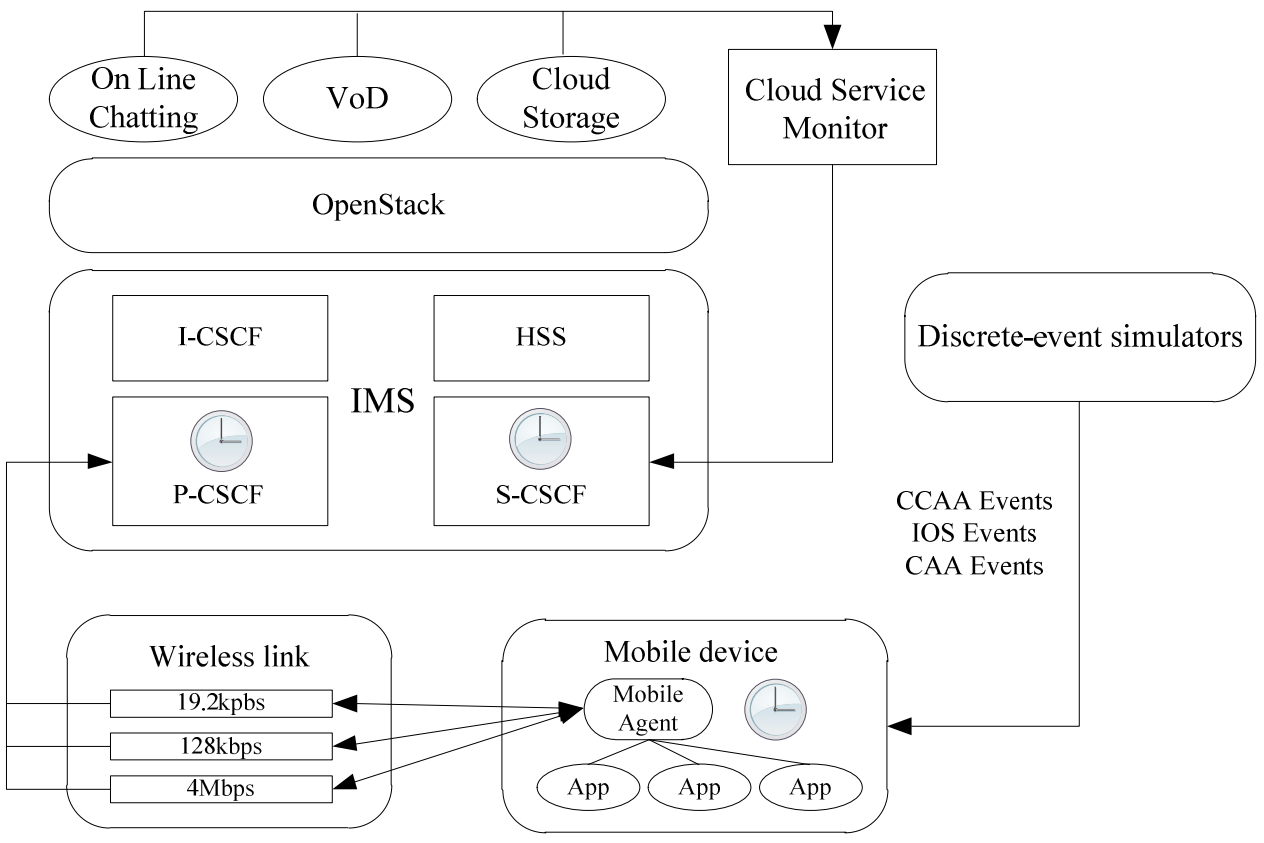

Fig. 5 Simulation Environment

To verify our analytic model, we have developed a Java-based test bed, depicted in Fig. 5 which includes several discrete-event simulators, the IMS network entities including P-CSCF, I-CSCF, S-CSCF and HSS, cloud services and a simple mobile device module. A timer is contained in the mobile device, P-CSCF and S-CSCF. The discrete-event simulators simulate the mobility behaviors of mobile device by generating user movement, session initiation and cloud service accessing events, which all follows exponential distribution. IMS network entities, Cloud Service Monitor and the mobile agent are developed based on the open source code SIP stack SIPp [19], which realized the basic transaction process. We use the OpenStack to provide three different vitalized machines, and develop the on line chatting, VoD and cloud storage on the cloud. Finally, the wireless link layer is simulated by several queues which carry some background traffic and communicate with mobile device with the specified packet loss rate.

The power cost of a SIP message is set to be the length of the message. For example the power cost for transmitting a re-registration message (225 bytes) is 225 W, where W is a constant of proportionality for the power consumed. The power cost for a mobile device update its timer cloud service is set to 10 W. Table 1 lists the parameters used in our simulations.

Table 1 Parameter values in Simulation

| Parameter | value |
|---|---|
| REGISTER | 225 bytes |
| INVITE | 810 bytes |
| 200OK | 100 bytes |
| ACK | 60 bytes |

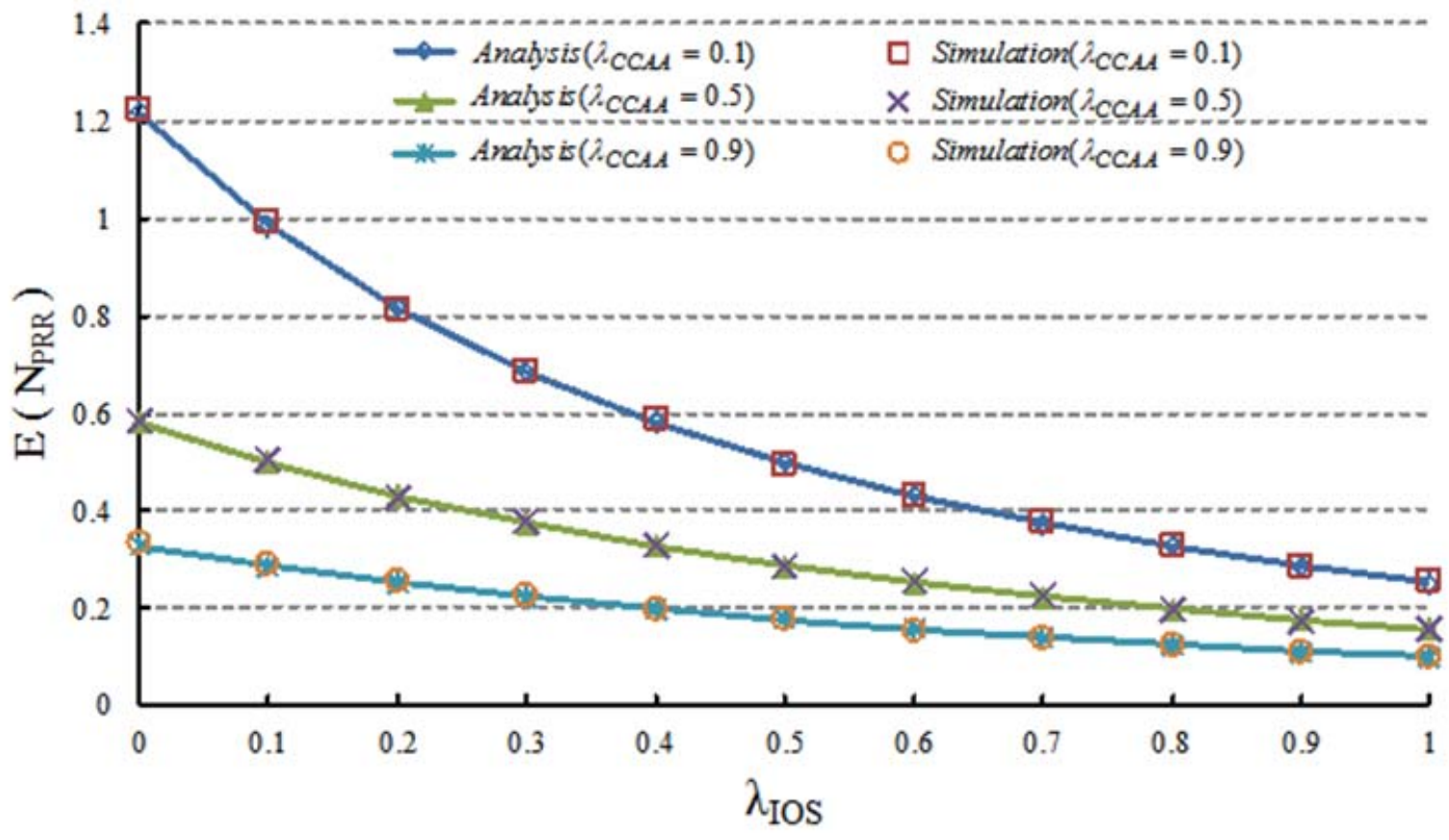

Fig. 6 The simulation and analytical value of PRR event number with $\lambda_{IOS}$

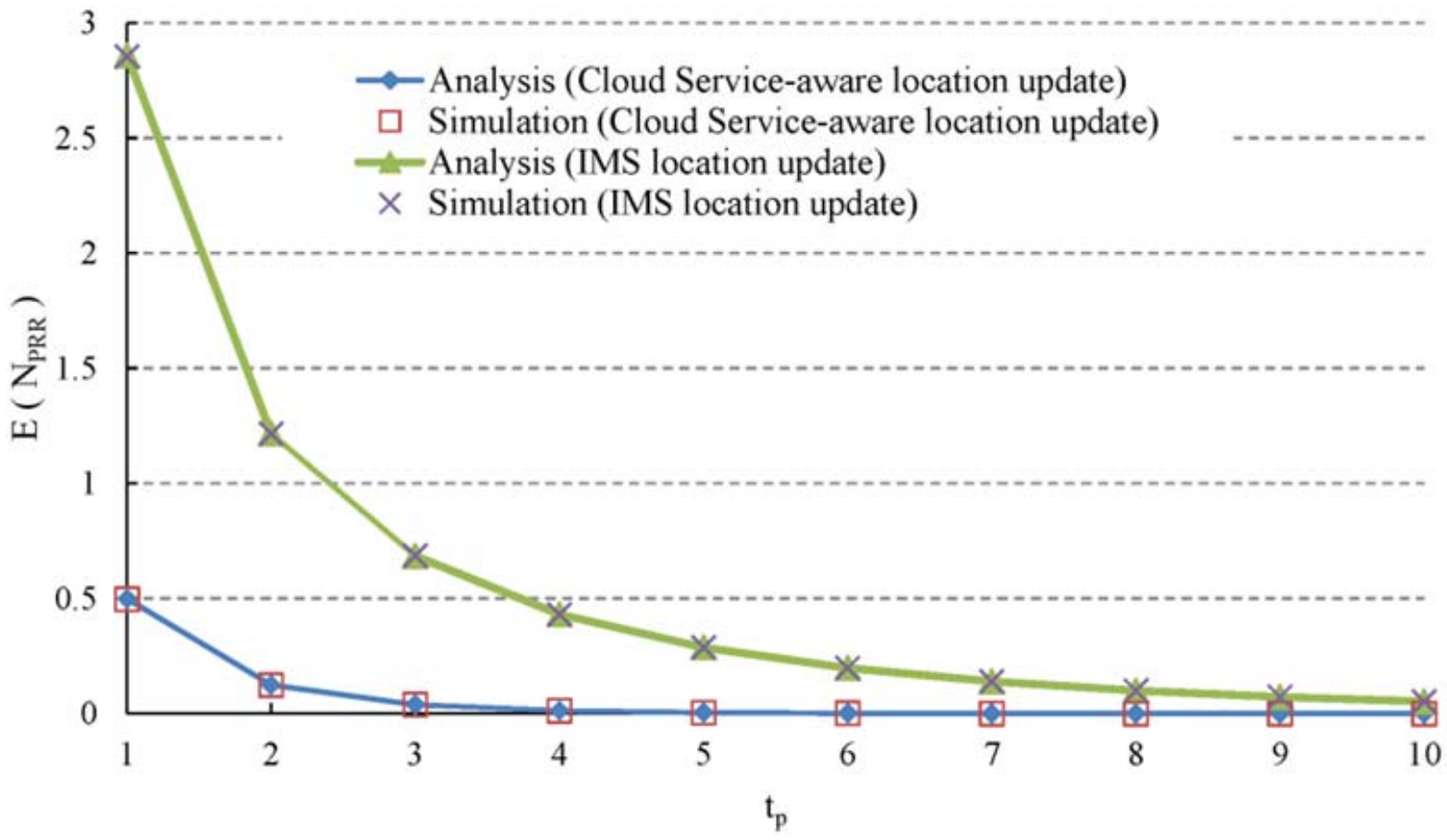

Fig. 7 The simulation and analytical value of PRR event number with $t_p$

From the analysis in Section III, the power cost for location update is decided by the number of periodic registration for refreshing the timer. Fig. 6 and Fig. 7 depict the number of periodically registration from simulation and analytical model. With different session arrival rates ($\lambda_{IOS}$) and timer lengths ($t_p$), the simulation and analytical results are almost the same. The errors are due to the number of random generated discrete-events. For example, if several more handoff events generated during the simulation period, the simulation result is bigger than the analytical result, and vice versa. As the error rate is under 1%, these experiments have verified that the analytic model is consistent with the simulation results. From Fig. 6 and Fig. 7, we can also see that the number of periodic registration for cloud service-aware location update is less than that for the IMS location update. By introducing the session establishment and cloud service accessing events, cloud service-aware location update reduces the number of periodic update events but does not impact the location update and mobile device's re-registration.

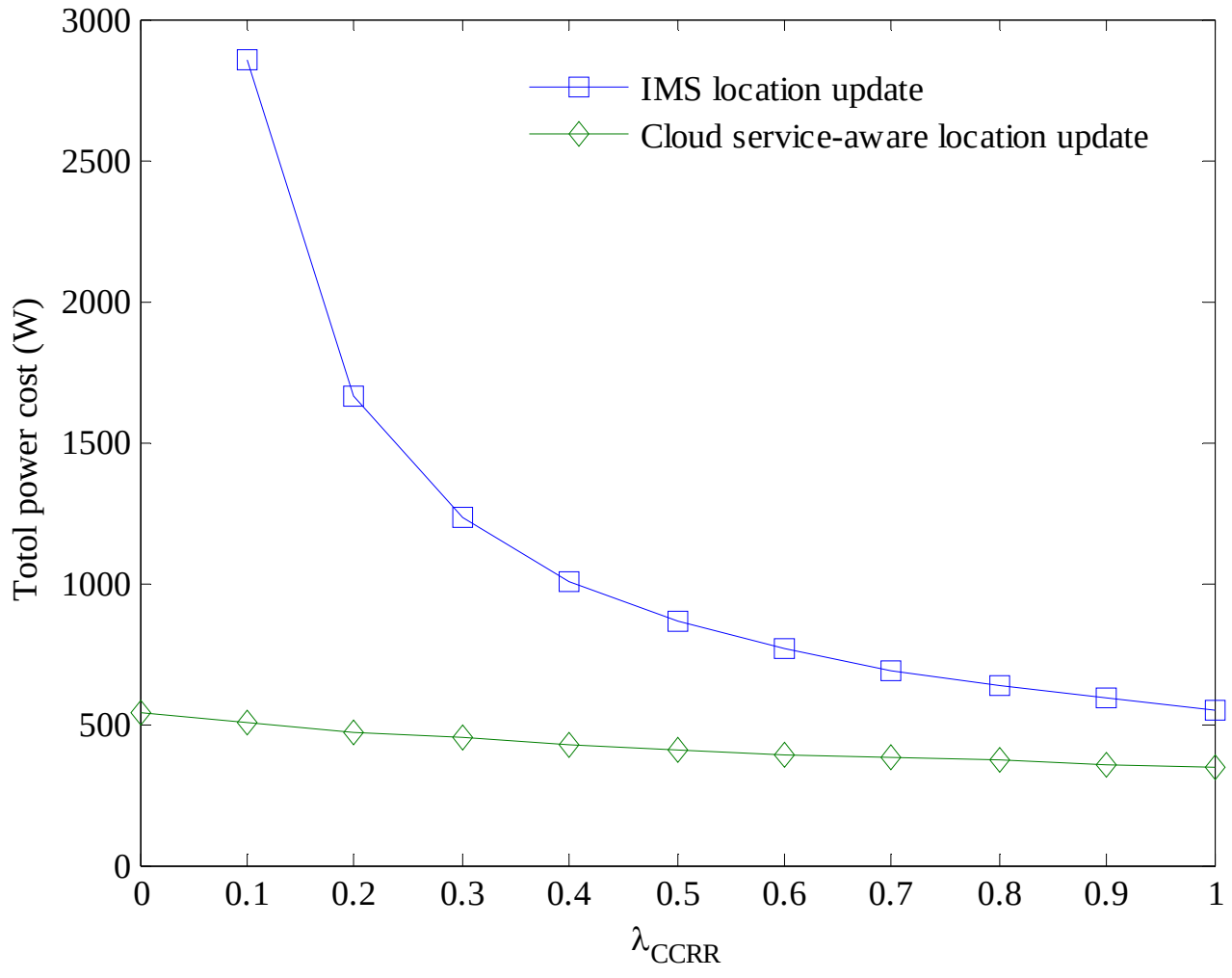

Fig. 8 The power cost for IMS and cloud service-aware location update with $\lambda_{CCRR}$ . $t_p = 1$ , $\lambda_{IOS}$ =0.5, $\lambda_{CSA}$ =0.5.

Fig. 8 is the comparing result of IMS and cloud service-aware location update with various arrival rate of CCRR event. We can see that introducing the IOS and CSA event to help update the mobile device's location, the power cost for cloud service-aware location update with dynamic location update is obviously smaller than that for IMS location update with periodic location update. When the arrival rate of CCRR event increases, the power cost for both the IMS and cloud service-aware location update decreases. The reason is the CCRR event can refresh the timer in IMS, which puts off the PRR event. Then, within a period of time, the number of PRR event decreases and the consumed energy of the mobile device is saved.

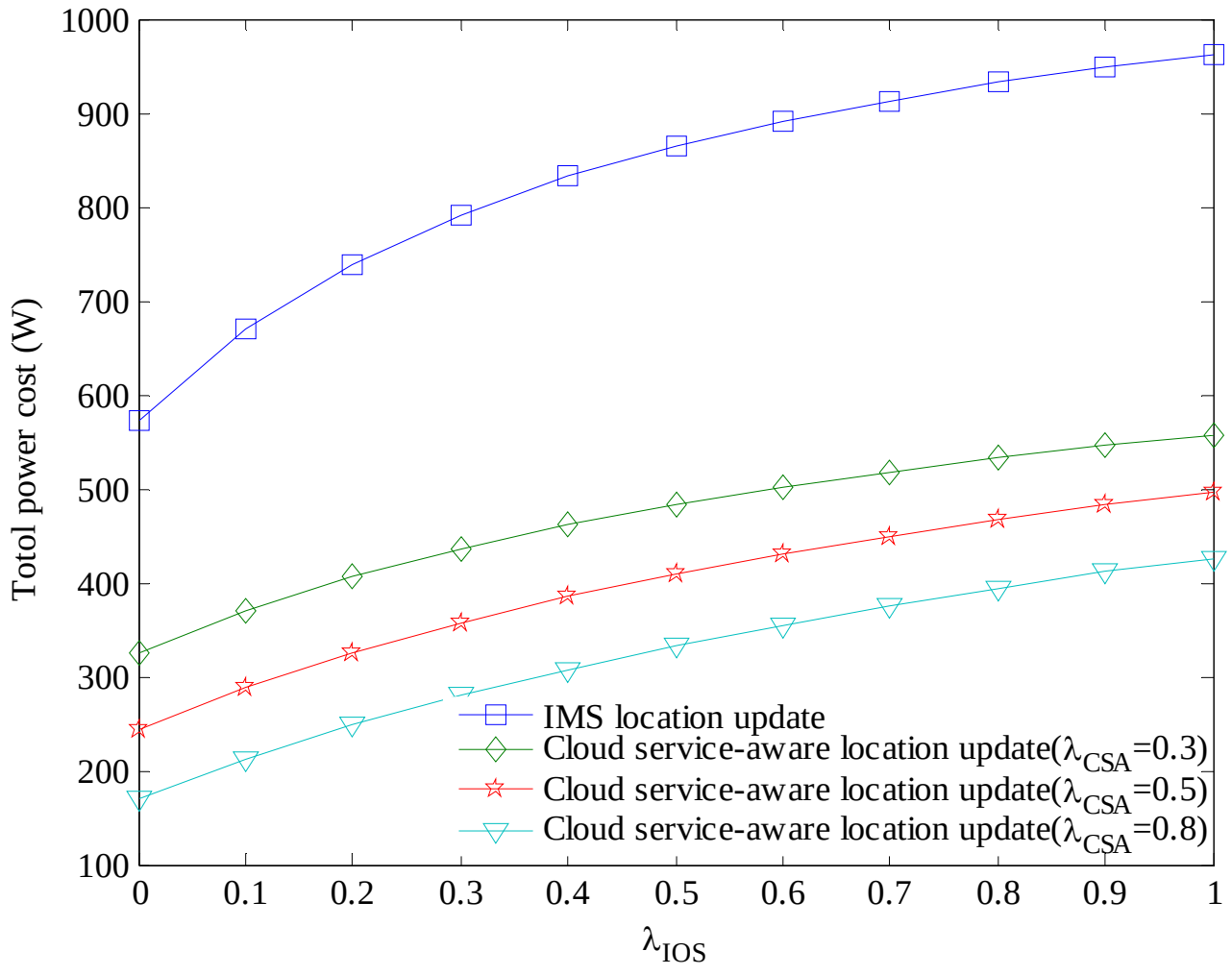

Fig. 9 Total cost for IMS and cloud service-aware location update with $\lambda_{IOS}$ . $t_p = 1$ and $\lambda_{CCRR}$ =0.5.

Fig. 9 depicts the power cost for the IMS and cloud service-aware location update under the condition of different IOS event arrival rates. When the IOS event arrival rate increases, both the power cost for IMS and cloud service-aware location update increase. This is because the energy for processing the session incoming and terminating increases, which is contained in the total calculated power cost. In Fig. 9, the power costs for cloud service-aware location update are all lower than that for IMS location update. And by the increasing of the arrival rate of cloud service accessing, the power cost for cloud service-aware location update is decreasing. The reason is by perceiving the mobile user's behavior and taking use of CSA event, the timer in IMS can be refreshed without periodic registration update.

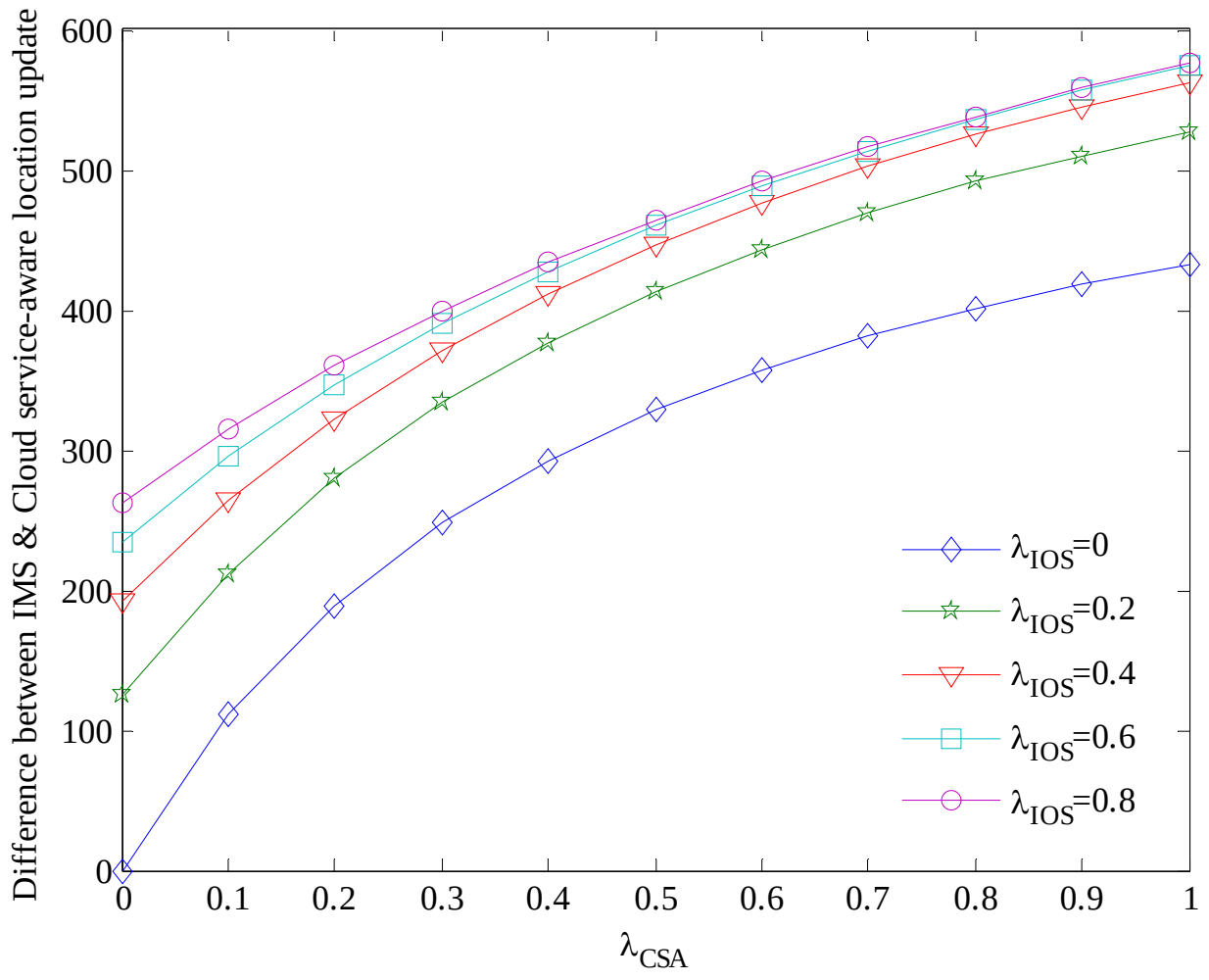

Fig. 10 The power cost between IMS and Cloud service-aware location update with $\lambda_{CSA}$. $t_p = 1$ and $\lambda_{CCRR}$ =0.5

Fig. 10 shows the difference of energy consumed between periodic location update in IMS and dynamic location update in new mechanism. The difference value increases when the arrival rate of CSA event increasing. The reason is cloud service-aware location update takes use of the cloud service accessing event to refresh timer, which reduces the number of PRR events and its power consumption. When $\lambda_{CCA} = 0$ and $\lambda_{IOS} = 0$, there is no session or cloud service event for update, and the timer refreshing in IMS is only based on the CCRR and PRR events. In this case, the IMS location update has the same performance with cloud service-aware location update, and the power consumptions of the mobile device of the two mechanisms are same. When the arrival rate of IOS event grows, the difference value becomes high. Because the cloud service-aware location update can refresh timer without the CCRR and PRR events, but the IMS location update need to use the PRR events, which results that the power cost for cloud service-aware location update is much less than that for IMS location update.

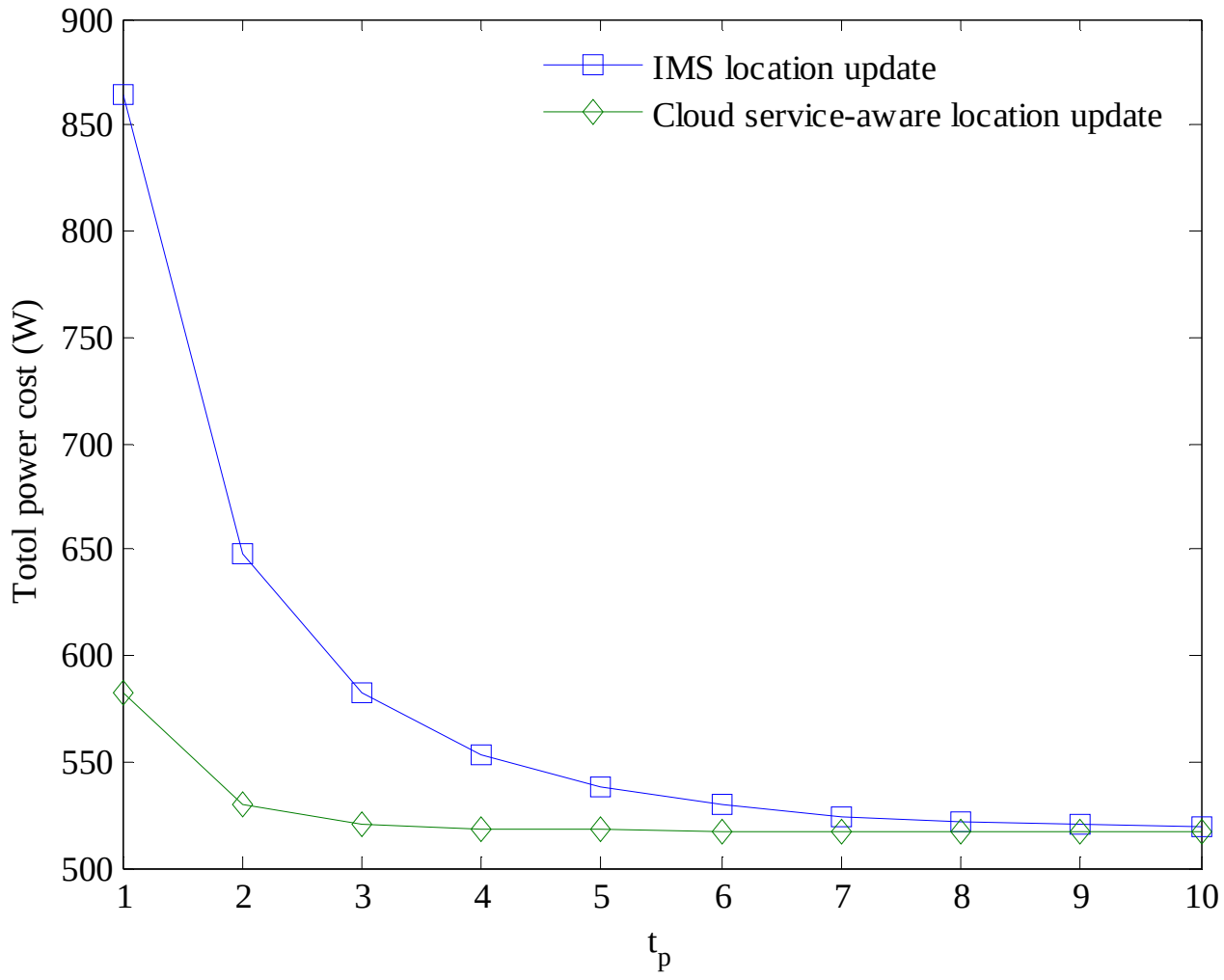

Fig. 11 The power cost between IMS and Cloud service-aware location update with $t_p$. $\lambda_{CCRR} = \lambda_{IOS} = \lambda_{CSA} = 0.3$

Fig. 11 depicts the total power cost with various register timer values. Along with timer value increasing, the update interval becomes large, then the times of periodic update decreases, which results in the power cost for both the IMS and cloud service-aware location update decreasing. But in this case, the power cost for cloud service-aware location update is smaller than that for IMS location update.

## VI. Conclusion

In this article, we focus on the mobile cloud computing where the mobile device acts like a thin client connecting to the remote server through wireless network. IMS provides an IP-based control plan for the mobile cloud computing environment, especially

QoS guarantee and mobility management. But considering the features of cloud computing services and the limited battery life of the mobile devices, IMS location update is not efficient for the energy consumption. By perceiving the cloud service that the mobile users accessing, we proposes an optimized location update techniques that support user mobility effectively. The service-aware location update detects the presence and location of the mobile device by the mobile users' behaviors such as access network change, session establishment, and cloud service accessing. The reduction of periodic registration makes energy consumption of mobile devices for location updates less. The evaluation results demonstrate that the service-aware location update minimizes unnecessary energy consumption for the location update.

## REFERENCES

[1] Niroshinie Fernando, Seng Wai Loke, Wenny Rahayu. Mobile cloud computing: A survey. Future Generation Comp. Syst.; vol. 29, pp.84-106 (2013)

[2] 3GPP TS 23.228, V.8.2.0. IP Multimedia System (IMS), 2007

[3] Hoang T. Dinh, Chonho Lee, Dusit Niyato, and Ping Wang, A Survey of Mobile Cloud Computing Architecture, Applications, and Approaches. Wireless Communications and Mobile Computing, DOI: 10.1002/wcm.1203 , 2011

[4] Paolo Bellavista, Giuseppe Carella, Luca Foschini, Thomas Magedanz, Florian Schreiner, Konrad Campowsky. QoS-aware elastic cloud brokering for IMS infrastructures. IEEE ISCC 2012, 157-160

[5] M. Garcia-Martin and G. Camarillo. Multiple-recipient MESSAGE requests in the session initiation protocol (SIP). IETF RFC 5365, 2008

[6] Pieter Simoens, Filip De Turck, Bart Dhoedt, Piet Demeester. Remote Display Solutions for Mobile Cloud Computing. IEEE Computer, 44 (2011) 46-53.

[7] Weixiong Rao, Kai Zhao, Eemil Lagerspetz, Pan Hui, Sasu Tarkoma, Energy-Aware Keyword Search on Mobile Phones. 2012 Sigcomm workshop on Mobile cloud computing,59-64

[8] Fangming Liu, Peng Shu. eTime: Energy-Efficient Mobile Cloud Computing for Rich-Media Applications. IEEE COMSOC MMTC E-Letter, 8, 2013.

[9] Seonggeun Ryu, Kyunghye Lee, Youngsong Mun, Optimized fast handover scheme in Mobile IPv6 networks to support mobile users for cloud computing. The Journal of Supercomputing, 59 (2012) 658-675

[10] Min Choi, Jonghyuk Park, Young-Sik Jeong, Mobile cloud computing framework for a pervasive and ubiquitous environment, The Journal of Supercomputing, 64 (2013) 331-356

[11] Ayush Vrat, Manish Sachan, Aarti Gautam Dinker, Deepanshu Arora, Anurika Vaish, S. Venkatesan, Performance Analysis of Enhanced Mobility Model in Cloud Computing, Recent Trends in Information Technology (ICRTIT), 2011, 638-643

[12] Yi-Bing Lin, Per-Chun Lee, Imrich Chlamtac. Dynamic Periodic Location Area Update in Mobile Networks. IEEE Transactions on Vehicular Technology, 51 (2002) 1494-1501

[13] Yang Xiao, Hui Chen. Optimal periodic location area update for mobile telecommunications networks. IEEE Transactions on Wireless Communications, 5 (2006) 930-937.

[14] Yang Xiao, Hui Chen. Periodic Location Area Update Schemes for UMTS 3G Mobile Networks: Optimality and Comparison. IEEE International Conference on Communications, 3(2006) 955-960.

[15] Yang Xiao, Hui Chen. An Analytical Model of the ODPLAU Scheme for Telecommunication Networks. IEEE Wireless Communications and Networking Conference 2007 (WCNC), 3194-3198

[16] Hasan Tuncer, Yoshihiro Nozaki, Nirmala Shenoy. Virtual mobility domains-A mobility architecture for the future Internet. IEEE ICC 2012, 2774-2779.

[17] Ronny Yongho Kim, Ritesh Kumar Kalle, Debabrata Das. Joint paging area and location update optimization for IEEE 802.16m idle mode. *Computer Networks*, 55 (2011) 3744-3758

[18] Lin, Y.B. Reducing Location Update Cost in a PCS Network. IEEE/ACM Transactions on Networking, 5 (1997) 25-33

[19] SIPp [OL], <http://sipp.sourceforge.net/> [Oct. 2011]